\documentclass[11pt,preprint,preprintnumbers,nofootinbib,superscriptaddress,amsmath,amssymb]{revtex4}
\usepackage{graphicx}
\usepackage{fancyhdr}
\usepackage{amsmath}
\usepackage{wrapfig}
\usepackage{color}

\begin{document}

\title{Doubly-charged Higgs bosons in the diboson decay scenario at the ILC\footnote{
This talk is based on the paper~\cite{KYY}, and the collaboration with Shinya Kanemura, Mariko Kikuchi, and Hiroshi Yokoya. }}

%

\author{Kei Yagyu}
\affiliation{Department of Physics, National Central University, Chungli 32001, Taiwan}

\begin{abstract}

The Higgs Triplet Model (HTM) is one of important examples for extended Higgs sectors, because 
tiny neutrino masses can be simply explained.   
Unlike the canonical type-I seesaw model, a scale of new particles can be taken as $\mathcal{O}(100)$ GeV keeping 
an enough amount of production cross section for direct searches at collider experiments. 
In the HTM, there appear doubly-charged Higgs bosons $H^{\pm\pm}$, and detection of them is a key to probe the model. 
The decay property of $H^{\pm\pm}$ depends on the magnitude of the vacuum expectation value of the triplet field $v_\Delta$. 
When $v_\Delta$ is smaller than about 1 MeV, $H^{\pm\pm}$ can mainly decay into the same-sign dilepton, and the lower mass limit for $H^{\pm\pm}$
had been taken to be about 400 GeV at the LHC. 
On the other hand, if $v_\Delta$ is larger than about 1 MeV, $H^{\pm\pm}$ can mainly decay into the same-sign diboson. 
In this case, the mass bound cannot be applied, so that the scenario based on light $H^{\pm\pm}$ is still possible. 
In this talk, we discuss the phenomenology of the same-sign diboson decay scenario of $H^{\pm\pm}$. 
First, we review the mass bound from the current collider experiments given in Ref.~\cite{KYY}.  
We then discuss the strategy for detection of $H^{\pm\pm}$ at the ILC.

\end{abstract}

\maketitle

\thispagestyle{fancy}


\section{Introduction}

The Higgs boson has been discovered at the LHC, 
and its properties are consistent with those of the Higgs boson in the Standard Model (SM)~\cite{Higgs_LHC}. 
Although the minimal Higgs sector assumed in the SM can explain this situation by the most economical way, 
we still do not know {\it what is the true structure of the Higgs sector}. 
In fact, in the Higgs sector with additional isospin scalar multiplets such as singlets, doublets and triplets, 
the discovered Higgs boson can be well explained as in the SM. 
Such a non-minimal Higgs sector often appears in a new physics model which can explain phenomena 
beyond the framework of the SM; e.g., neutrino oscillations, 
the existence of dark matter and baryon asymmetry of the Universe. 
Therefore, by the determination of the structure of Higgs sector from collider experiments, 
we can get a clue for new physics models. 

The type-II seesaw scenario~\cite{typeII} is one of important examples for new physics which deduces a non-minimal Higgs sector, where 
tiny neutrino masses can be simply explained. 
The Higgs sector in this scenario corresponds to the Higgs Triplet Model (HTM) which is composed of the 
isospin doublet scalar field $\Phi$ with the hypercharge $Y=1/2$ and the isospin triplet scalar field $\Delta$ with $Y=1$. 
From new Yukawa interactions $h_{ij}\overline{L_L^{ic}}(i\tau_2)\Delta L_L^j$ for the left-handed lepton doublet field $L_L$,   
Majorana masses are generated at the tree level as 
\begin{align}
(M_\nu)_{ij} = \sqrt{2}h_{ij}v_\Delta, \label{Eq:numass}
\end{align}
where $v_\Delta=\sqrt{2}\langle\Delta^0\rangle$ is the vacuum expectation value (VEV) of $\Delta$. 
The important point in Eq.~(\ref{Eq:numass}) is that 
the mass of triplet scalar field does not enter in this expression. 
Therefore, we can consider triplet scalar boson masses to be $\mathcal{O}(100)$ GeV.  
In such a case the HTM can be tested at collider experiments. 
 
In this talk, we focus on the direct detection of $H^{\pm\pm}$ at the LHC and at the ILC. 
There are three decay modes for $H^{\pm\pm}$, i.e., the same-sign dilepton decay ($H^{\pm\pm}\to \ell^\pm\ell^\pm$), 
the same-sign diboson decay ($H^{\pm\pm}\to W^{\pm(*)} W^{\pm(*)}$), and the cascade decay 
($H^{\pm\pm}\to H^\pm W^{\pm*}$), where $H^\pm$ are the singly-charged Higgs bosons mainly originated from the triplet field. 
Collider phenomenology of $H^{\pm\pm}$ with the same-sign dilepton, the same-sign diboson and the cascade decay has been 
studied in Refs.~\cite{dilepton,Han},~\cite{Han} and~\cite{cascade}, respectively. 
Among these three decay channels, when $H^{\pm\pm}$ mainly decay into the same-sign dilepton, 
there had been a lower limit on the mass of $H^{\pm\pm}$ $(m_{H^{++}})$ to be about 400 GeV from the LHC~\cite{400GeV_ATLAS,400GeV_CMS}. 
However, this mass bound cannot be applied to the case where $H^{\pm\pm}$ can mainly decay into the same-sign diboson
In this talk, we focus on the same-sign diboson decay scenario of $H^{\pm\pm}$. We first discuss the current bound on $m_{H^{++}}$ from 
the LEP and the LHC experiments. We then consider the phenomenology of a light $H^{\pm\pm}$ scenario at the ILC. 


\section{Important Features at the Tree Level}

There are several characteristic features in the HTM. 
First of all, the electroweak rho parameter deviates from unity at the tree level as
\begin{align}
\rho_\text{tree} =\frac{v_\phi^2+2v_\Delta^2}{v_\phi^2+4v_\Delta^2}\simeq 1-\frac{2v_\Delta^2}{v_\phi^2}, \label{Eq:rho}
\end{align}
where 
$v_\phi$ is the VEV of $\Phi$, and it satisfies $v^2=v_\phi^2+2v_\Delta^2\simeq (246~\text{GeV})^2$. 
The experimental value is give as $\rho_{\text{exp}}=1.0004^{+0.0003}_{-0.0004}$~\cite{PDG}, so that 
$v_\Delta$ is constrained to be smaller than about 3.5 GeV at the 95\% confidence level (CL) from Eq.~{\ref{Eq:rho}).  
Because of the smallness of $v_\Delta$, the mixing between $\Phi$ and $\Delta$ is very weak. 
Therefore, the component scalar fields in $\Delta$ are almost the mass eigenstates, where 
there are the doubly-charged $H^{\pm\pm}$, the singly-charged $H^\pm$, a CP-odd $A$ and a CP-even $H$ scalar bosons. 
We can call them triplet-like Higgs bosons. 

Second, there appear relationships among the masses of triplet-like Higgs bosons under $v_\Delta \ll v_\phi$; i.e., 
$m_{H^{++}}^2-m_{H^{+}}^2 \simeq m_{H^+}^2-m_A^2$ and $m_H^2 \simeq m_A^2$. 
From this relation, three patterns of the mass spectrum can be considered; namely, 
$m_{A}>m_{H^+}>m_{H^{++}}$, $m_{H^{++}}>m_{H^+}>m_A$ and all of them are degenerate in mass~\cite{cascade,HTM_mass_diff}. 
In the following discussion, we concentrate on the degenerate case, and we discuss the phenomenology of $H^{\pm\pm}$. 

\begin{figure}[t]
\begin{center}
 \includegraphics[width=80mm]{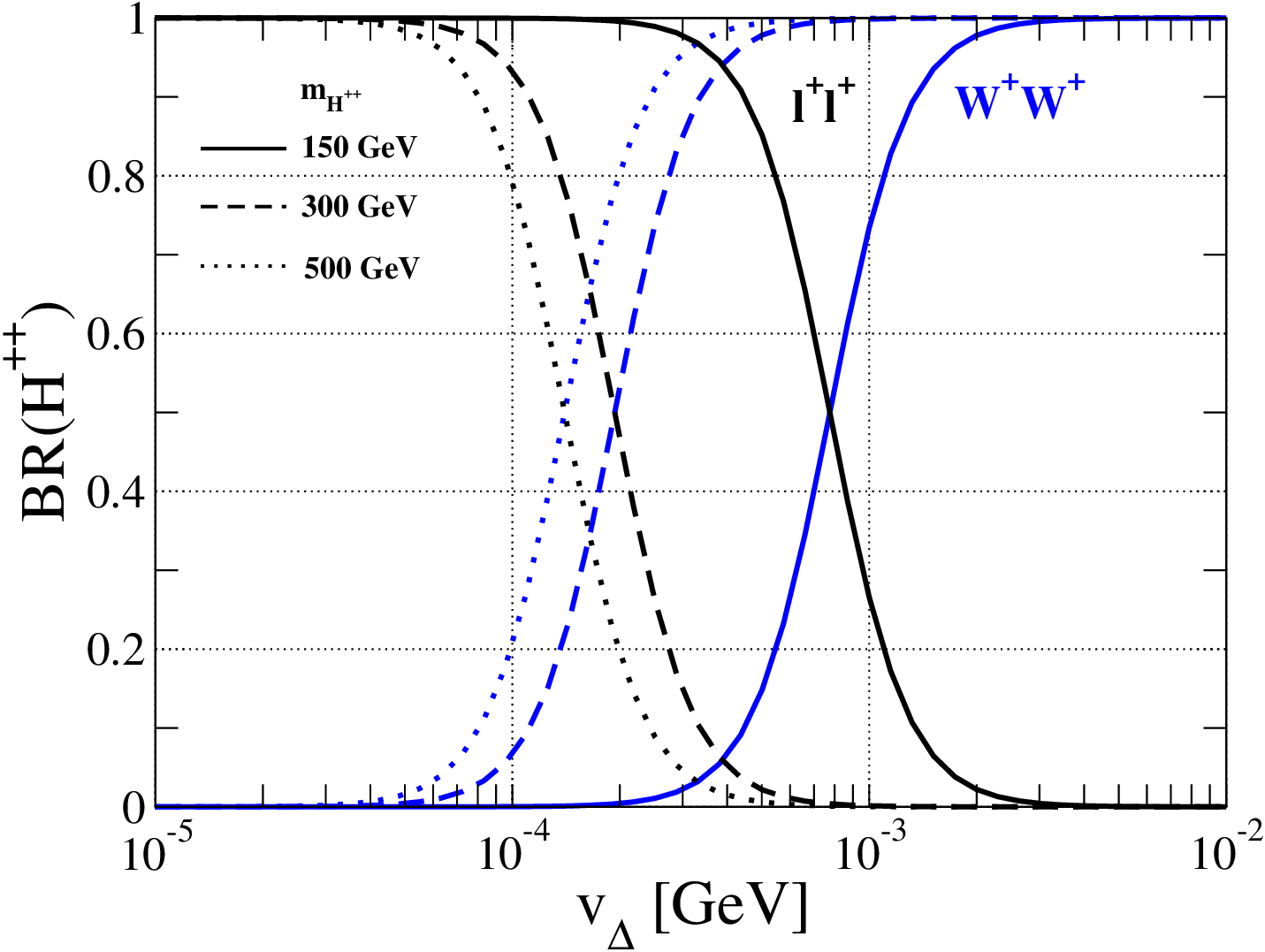}
   \caption{
Decay branching ratio of $H^{++}$ as a function of $v_\Delta$ with $m_{H^+}=m_{H^{++}}$. 
The solid, dashed and dotted curves respectively show the results in the case of $m_{H^{++}}=150$, 300 and 500 GeV. }
   \label{Fig:BR}
\end{center}
\end{figure}

Finally, 
the gauge interactions and Yukawa interactions of $H^{\pm\pm}$ 
are derived from the kinetic term and the neutrino Yukawa interactions as 
\begin{align}
\mathcal{L}_\text{int}=
-gm_W\frac{\sqrt{2}v_\Delta}{v}g_{\mu\nu}H^{++}W^-_\mu W^-_\nu
-\frac{(M_\nu)_{ij}}{\sqrt{2}v_\Delta}\overline{\ell^{ic}}P_L\ell^jH^{++} +\text{h.c.}, 
\end{align}
where $m_W$ is the W boson mass. 
Assuming all the elements in $(M_\nu)_{ij}$ to be 0.1 eV, the decay branching ratio of $H^{\pm\pm}$ is calculated as in Fig.~\ref{Fig:BR}. 
We can see that the dominant decay mode is changed from the same-sign dilepton mode to the same-sign diboson mode at $v_\Delta=0.1$-1 MeV. 

\section{Constraint on $m_{H^{++}}$ from Collider Experiments}

\begin{figure}[t]
\begin{center}
 \includegraphics[width=70mm]{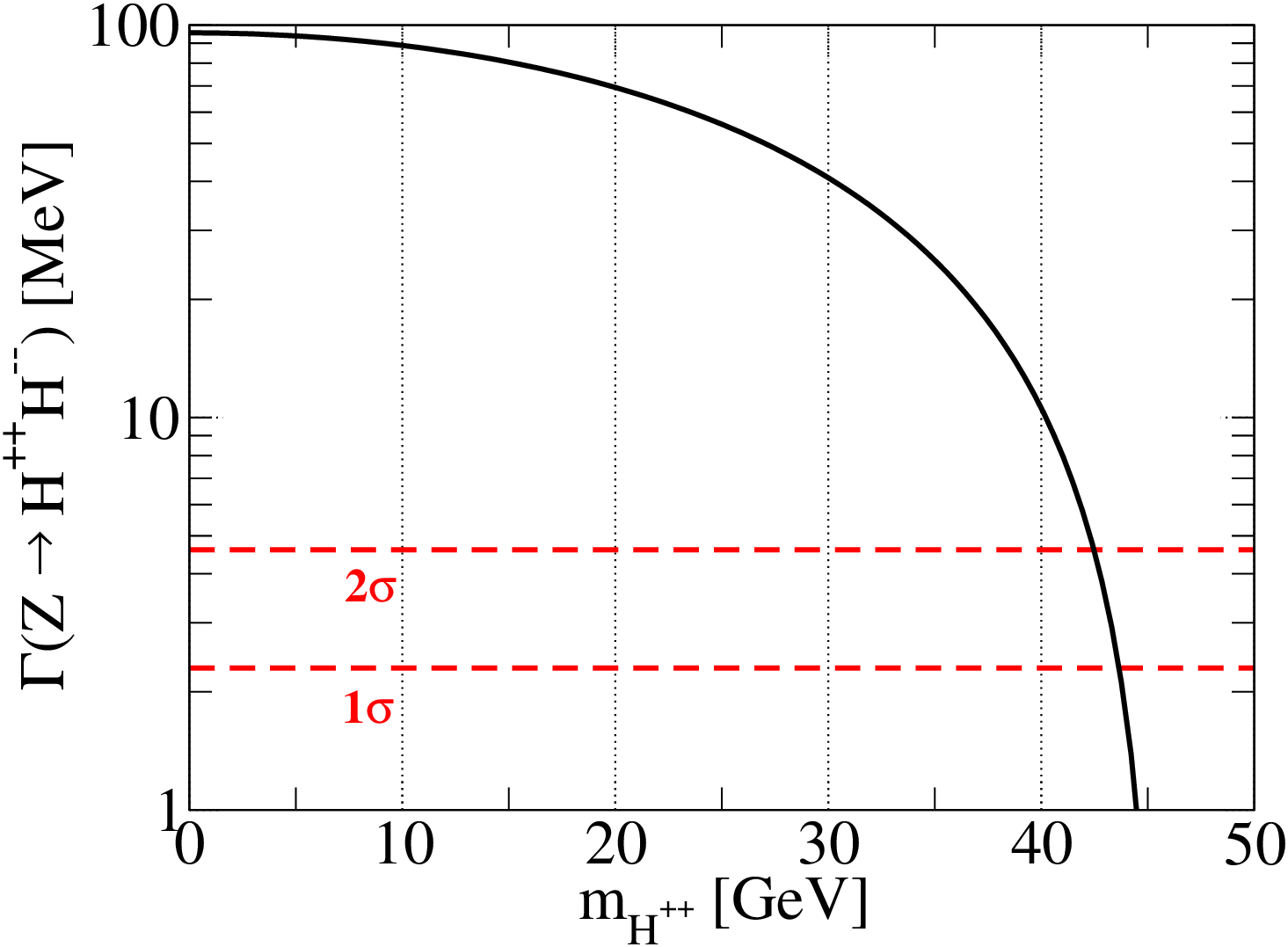}
   \caption{
Decay rate of $Z\to H^{++}H^{--}$ as a function of $m_{H^{++}}$. 
The 1$\sigma$ and 2$\sigma$ error bars of the measured Z boson width are also shown by the dashed horizontal lines.  }
   \label{Fig:Gamma_Z}
\end{center}
\end{figure}

At the LEP experiment, the width of the Z boson has been precisely measured as 
$\Gamma_Z(\rm exp)=2.4952\pm0.0023$~GeV~\cite{PDG}. 
If the mass of $H^{\pm\pm}$ is smaller than the half of the Z boson mass, 
the Z boson width can be significantly modified due to the 
$Z\to H^{++} H^{--}$ decay.  
In Fig.~\ref{Fig:Gamma_Z}, we show the decay rate of $Z\to H^{++}H^{--}$ as a function of $m_{H^{++}}$. 
From this figure, we obtain the lower bound on $m_{H^{++}}$ to be about 43 GeV at the 95\% CL. 
Because the bound is obtained only from the width of the Z boson, 
this constraint does not depend on the decay channel of $H^{\pm\pm}$. 

At the LHC, $H^{\pm\pm}$ are produced by the Drell-Yan process $pp\to Z^*/\gamma^* \to H^{++}H^{--}$ and 
the associated process $pp\to W^* \to H^{\pm\pm}H^{\mp}$. 
The search for $H^{\pm\pm}$ in the dilepton decay scenario has been performed at the LHC.  
The strongest lower limit on $m_{H^{++}}$ has been given by 459 GeV~\cite{400GeV_CMS} at the 95\% CL assuming the 100\% decay of 
$H^{\pm\pm}\to \mu^\pm \mu^\pm$ from the 7 TeV and 4.9 fb$^{-1}$ data. 
This bound becomes weaker as 395 GeV~\cite{400GeV_CMS} when we only use the pair production process. 
However, when $H^{\pm\pm}$ mainly decay into the same-sign diboson, this bound can no longer be applied. 

\begin{figure}[t]
\begin{center}
 \includegraphics[width=70mm]{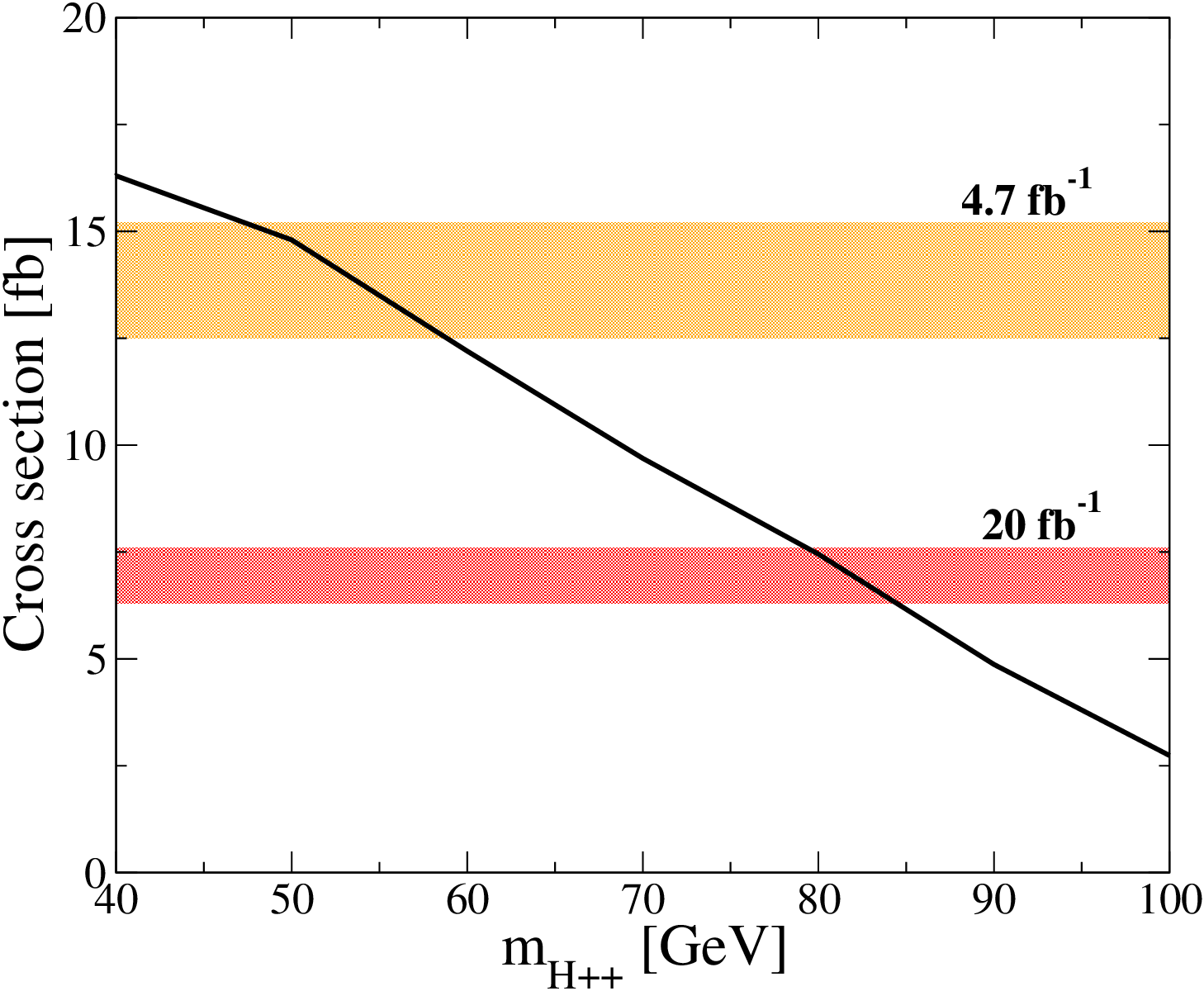}
\caption{The signal cross section expressed in Eq.~(\ref{Eq:signal}) as a function of $m_{H^{++}}$ with the collision energy to be 7 TeV from Ref.~\cite{KYY}. 
The light (dark) shaded band shows the 95\% CL (expected) upper bound for the cross section from the data with the integrate luminosity to be 4.7 fb$^{-1}$ (20 fb$^{-1}$). }
   \label{Fig:LHC}
\end{center}
\end{figure}

In Ref.~\cite{KYY}, the lower bound on $m_{H^{++}}$ has been taken by using the same-sign dilepton event measured at the LHC with 
7 TeV and 4.7 fb$^{-1}$ data~\cite{ATLAS_SS}. 
In Fig.~\ref{Fig:LHC}, the sum of the cross sections of 
\begin{align}
pp\to H^{++}H^{--}\to W^{+(*)}W^{+(*)}H^{--}\to \mu^+\mu^+ E_{\text{miss}}H^{--},\notag\\
pp\to H^{++}H^{-}\to W^{+(*)}W^{+(*)}H^{-}\to \mu^+\mu^+ E_{\text{miss}}H^{-}, \label{Eq:signal}
\end{align}
processes are shown as a function of $m_{H^{++}}$ in the case of $m_{H^+}=m_{H^{++}}$. 
It is seen that $m_{H^{++}}$ smaller than about 60 GeV is excluded at the 95\% CL. 
By the extrapolation of the data to 20~fb$^{-1}$ with the same collision energy, 
the lower limit is given to be 85 GeV.  
Therefore, a light $H^{\pm\pm}$ such as around 100 GeV is still allowed by the current data at the LHC.

\section{Detection of $H^{\pm\pm}$ at the ILC}

In this section, we discuss the detection of $H^{\pm\pm}$ in the diboson decay scenarios at the ILC. 
We first classify possible three scenarios which are
expected after the 300 fb$^{-1}$ data will be accumulated at the LHC with the 14 TeV energy as follows
\begin{enumerate}
\item $H^{\pm\pm}$ will be discovered at the LHC,
\item $H^{\pm\pm}$ will not be discovered at the LHC, and its mass bound  is smaller than $\sqrt{s}/2$,
\item $H^{\pm\pm}$ will not be discovered at the LHC, and its mass bound  is larger than $\sqrt{s}/2$,
\end{enumerate}
where $\sqrt{s}$ is the center of mass energy at the ILC. 
Case 1 is the most attractive scenario for testing the HTM, where $m_{H^{++}}$ would be measured at the LHC. 
Therefore, by focusing on the collision energy to be the half of $m_{H^{++}}$, 
precise measurements of the properties of $H^{\pm\pm}$ such as the mass, width and decay branching ratios are possible. 
In addition, loop effects of $H^{\pm\pm}$ to the Higgs boson couplings can be calculated by fixing $m_{H^{++}}$, 
and then we can compare the predictions with the precisely observed values. 
This can be the consistency check for measured $H^{\pm\pm}$. 
When Case 2 is realized, we can use the $e^+e^-\to H^{++}H^{--}$ production as the discovery mode of $H^{\pm\pm}$.   
As the indirect search, we can calculate deviation in Higgs boson couplings by fixing $m_{H^{++}}$ to be larger than 
the lower bound given from the LHC data. 
If Case 3 is realized, only the indirect search can be used to test the HTM. 

\begin{figure}[t]
\begin{center}
 \includegraphics[width=80mm]{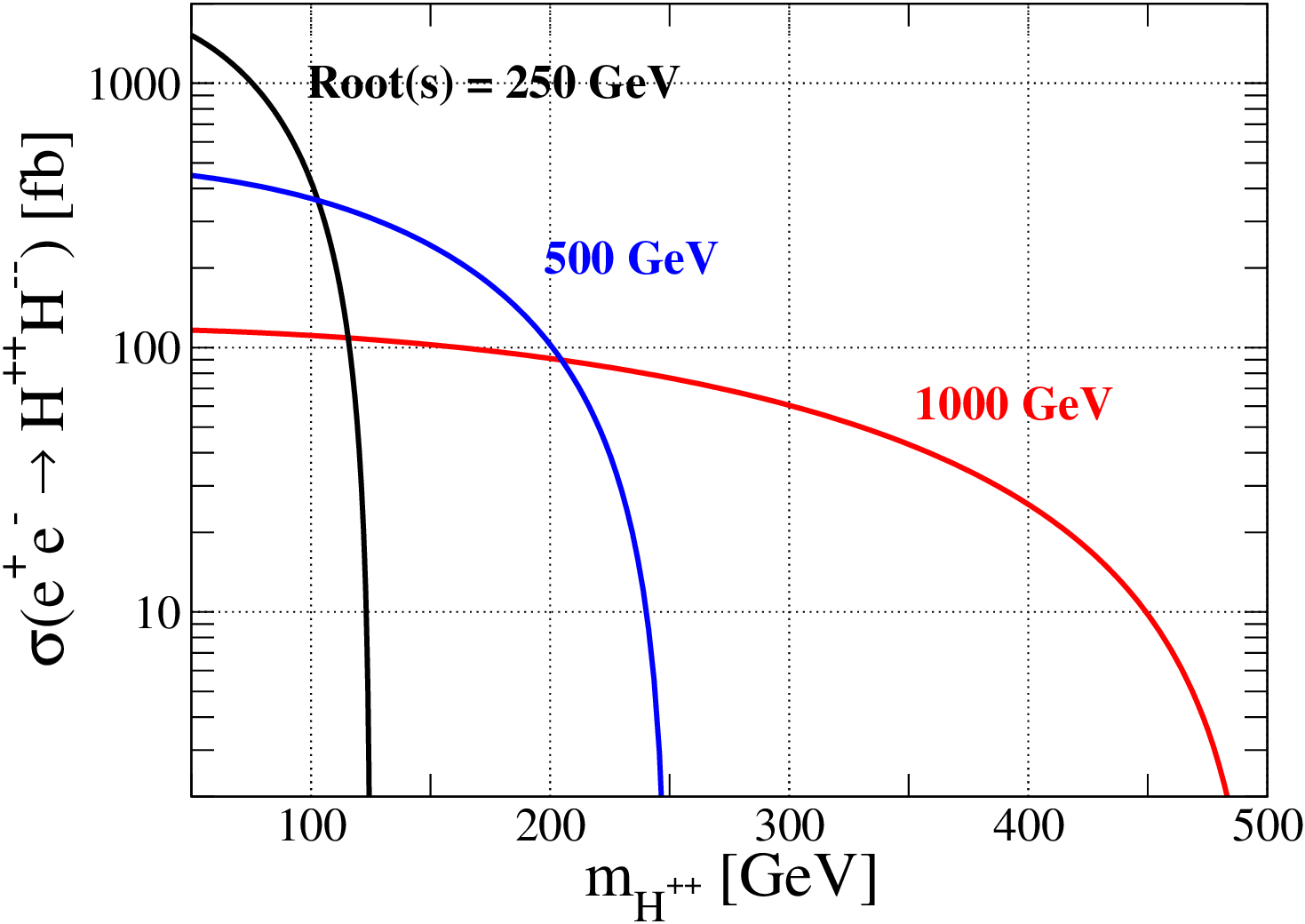}
   \caption{
Production cross section of the $e^+e^-\to H^{++}H^{--}$ process as a function of $m_{H^{++}}$. 
The black, blue and red curves are respectively the results with the collision energy $\sqrt{s}=$250, 500 and 1000 GeV.  }
   \label{Fig:sigma_Hpp}
\end{center}
\end{figure}

Let us suppose that Case 1 or Case 2 is realized. 
The production cross section of the $e^+e^-\to \gamma^*/Z^*\to H^{++}H^{--}$ process is given at the leading order by 
\begin{align}
\sigma(e^+e^-\to H^{++}H^{--})= \frac{\pi\alpha^2}{3s}\beta^3 (x_{H^{++}})
 \left[
 4Q_e^2
 + \frac{4v_eQ_e}{1-x_Z}
 \frac{1-2s_W^2}{s_W^2c_W^2}
 + \frac{v_e^2+a_e^2}{(1-x_Z)^2 }
\frac{(1-2s_W^2)^2}{s_W^4c_W^4} \right], 
\end{align}
where $\beta(x)=\sqrt{1-4x^2}$, $x_i=m_i^2/s$, $v_e=I_e^3/2-s_W^2Q_e$, $a_e=I_e^3/2$, $c_W=\cos\theta_W$ and $s_W=\sin\theta_W$ with 
$I_e^3$, $Q_e$ and $\theta_W$ being the isospin of electron, the electric charge of electron and weak mixing angle, respectively.  
In Fig.~\ref{Fig:sigma_Hpp}, the pair production cross section is shown as a function of $m_{H^{++}}$ in the cases with $\sqrt{s}=250$, 500 and 1000 GeV. 

We consider the signal and background events for the $e^+e^-\to H^{++}H^{--}$ process. 
In the diboson decay scenario, we expect the final state with the same-sign dilepton, missing energy and multi-jets; i.e., 
$e^+e^-\to H^{++}H^{--}\to \ell^+\ell^+ E_\text{miss}jjjj$, where $\ell=e,\mu$. 
The background comes from the four W bosons production; $e^+e^-\to W^+W^+W^-W^-\to \ell^\pm\ell^\pm E_\text{miss}jjjj$. 
When we take $\sqrt{s}=500$ GeV and the $m_{H^{++}}=230$ GeV as an example, 
we get the signal (background) cross section of $\ell^\pm\ell^\pm E_\text{miss}4j$
final state to be 1.07 fb (2.37$\times 10^{-3}$ fb) by using MadGraph5~\cite{MG5}. 
The above numbers are obtained after taking the following basic kinematic cuts 
\begin{align}
 p_T^\ell \geq 15 ~\text{GeV},\quad |\eta^\ell| \leq  2.5, 
\end{align}
where $p_T^\ell$ and $\eta^\ell$ are the transverse momentum and pseudo rapidity for $\ell$, respectively. 
Therefore, this process is almost background free. 

\begin{figure}[t]
\begin{center}
 \includegraphics[width=60mm]{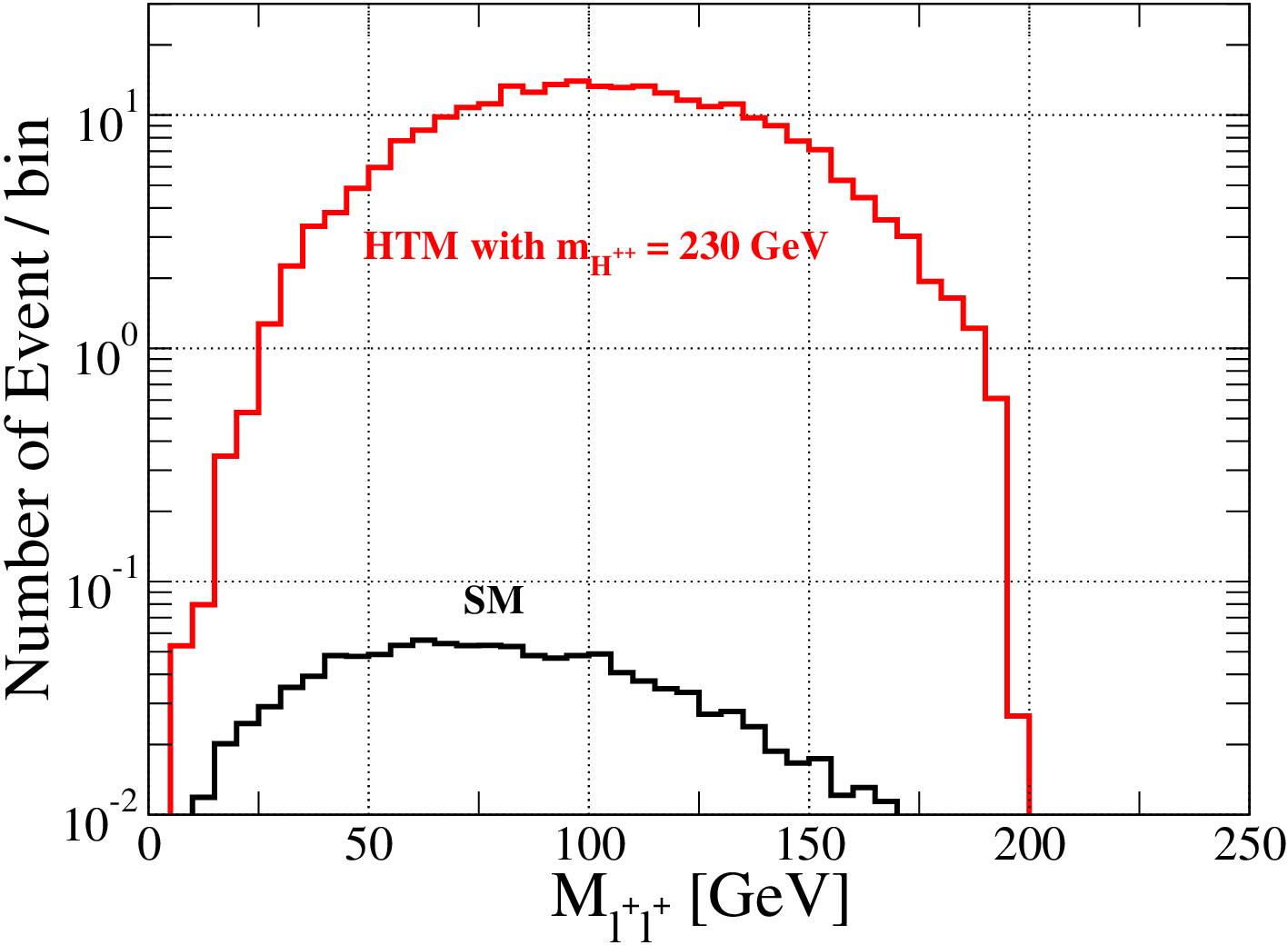}\hspace{5mm}
 \includegraphics[width=60mm]{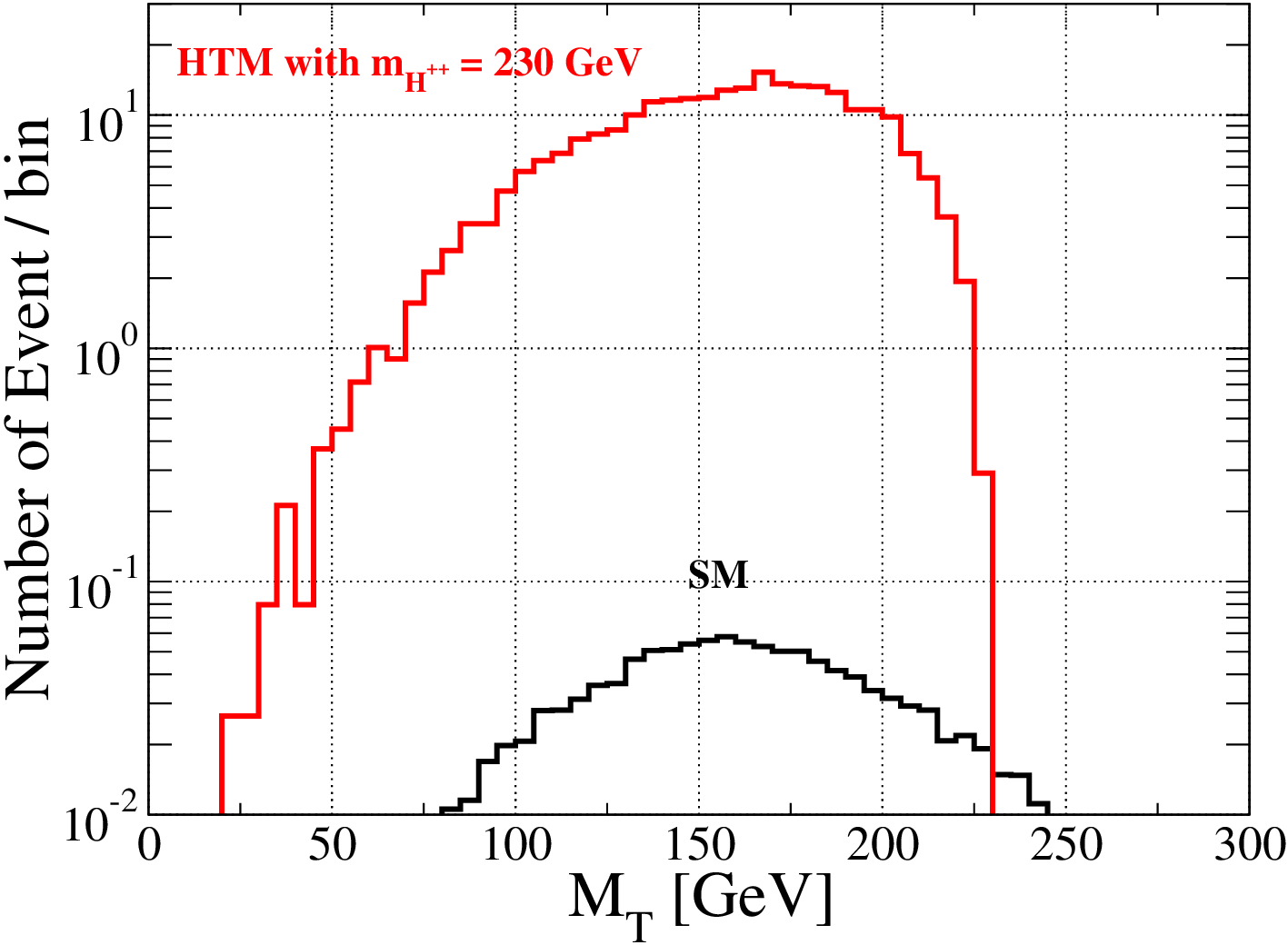}
   \caption{
The invariant mass distribution (left panel) and the transverse mass distribution (right panel) for the $\ell^+\ell^+$ and $\ell^+\ell^+E_{\text{miss}}$ systems, respectively, in the case of $m_{H^{++}}=230$ GeV and $\sqrt{s}=500$ GeV. The integrated luminosity is assumed to be 500 fb$^{-1}$.  }
   \label{Fig:distr}
\end{center}
\end{figure}

Fig.~\ref{Fig:distr} shows the invariant mass $M_{\ell^+\ell^+}$ for the $\ell^+\ell^+$ system (left panel) and the transverse mass $M_T$ (right panel) distributions for $\ell^+\ell^+E_{\text{miss}}$ system.
The red and black curves denote the distribution from the signal and background, respectively. 
There is an end point in the $M_T$ distribution at around 230 GeV which corresponds to $m_{H^{++}}$. 
Therefore, the $M_T$ distribution is useful to measure $m_{H^{++}}$. 

Finally, we would like to comment on the indirect search for the HTM from the precision measurements of the Higgs boson coupling constants. 
At the ILC, the Higgs boson couplings are expected to be precisely measured. 
For example, the Higgs boson couplings with the weak gauge bosons ($hZZ$ and $hWW$) 
and the Yukawa couplings ($hb\bar{b}$, $h\tau\bar{\tau}$ and $ht\bar{t}$) are expected to be measured with $\mathcal{O}(1)\%$ accuracy~\cite{ILC_White,ILC_TDR}. 
In the HTM, the loop induced $h\gamma\gamma$ coupling has been calculated in Refs.~\cite{hgg_HTM,KY,AKKY_full}. 
The one-loop corrections to the $hWW$, $hZZ$ and $hhh$ vertices have also been calculated in Refs.~\cite{AKKY,AKKY_full}. 
According to Ref.~\cite{AKKY_full}, it has been found that there is a correlation among the deviation in the Higgs boson couplings. 
For example, 
when the decay rate of $h\to \gamma\gamma$ deviates by 30\% (40\%)
from the SM prediction, deviations in the one-loop corrected $hVV$ and $hhh$ vertices are
predicted to be about $-0.1\%$ ($-2\%$) and $-10\%$ ($150\%$), respectively.
By comparing these deviations with the precisely measured value at the ILC, we can discriminate the HTM from the other models. 

\section{Conclusion}

We have discussed how we can test the HTM at collider experiments. 
The detection of $H^{\pm\pm}$ can be a direct evidence for the HTM, so that we have focused on the direct search for $H^{\pm\pm}$. 
The collider phenomenology of $H^{\pm\pm}$ can be drastically different depending on the main decay mode of $H^{\pm\pm}$. 
When $v_\Delta$ is smaller (larger) than about 1 MeV, $H^{\pm\pm}$ can mainly decay into the same-sign dilepton (diboson). 
If the same-sign dilepton mode is dominate, the lower mass bound of $H^{\pm\pm}$ has been taken to be about 400 GeV. 
However, this bound cannot be applied when the same-sign diboson decay is dominate. 

We then have studied the bound of $m_{H^{++}}$ in the same-sign diboson decay scenario. 
It has been found in Ref.~\cite{KYY} that the lower limit on the mass to be about 60 GeV can be obtained by using the same-sign dilepton data collected at the LHC with 7 TeV and 4.7 fb$^{-1}$.  
Therefore, a light $H^{\pm\pm}$ scenario is still possible in the same-sign diboson decay scenario. 

We have simulated the process $e^+e^- \to H^{++}H^{--}\to W^+W^+W^-W^-\to \ell^+\ell^+E_\text{miss}4j $ at the ILC. 
When we take $m_{H^{++}}=230$ GeV and 500 GeV for the collision energy, the signal cross section is about 1.1 fb. 
On the other hand, 
the corresponding background cross section from the four W bosons production is about 2.4$\times 10^{-3}$ fb, so that 
this process can be regarded almost background free. 
We have found that by looking at the end point in the transverse mass distribution of the same-sign dilepton plus missing system, 
we may be able to reconstruct the mass of $H^{\pm\pm}$.

\bigskip 
{\it Acknowledgments}

I would like to thank Shinya Kanemura, Mariko Kikuchi and Hiroshi Yokoya for fruitful collaborations. 
This work was supported in part by the National Science Council of R.O.C. under Grant No. NSC-101-2811-M-008-014.

\bigskip 

\end{document}